\documentclass[10pt,aip,jap,showpacs,floatfix,raggedbottom,floats,reprint,twocolumn]{revtex4-1}
\usepackage{graphicx,latexsym}
\usepackage{dcolumn}
\usepackage{amssymb,amsmath,bm}
\usepackage[hidelinks=true]{hyperref}
\usepackage{natbib}
\hypersetup{
  colorlinks   = true, %Colours links instead of ugly boxes
  urlcolor     = blue, %Colour for external hyperlinks
  linkcolor    = blue, %Colour of internal links
  citecolor    = red %Colour of citations
}

\begin{document}

%-----------------------------------------------------------------
\title{Coexisting spin and Rabi-oscillations at intermediate time\\ in electron transport 
       through a photon cavity}

\author{Vidar Gudmundsson}
\email{vidar@hi.is}
\author{Hallmann Gestsson}
\affiliation{Science Institute, University of Iceland, Dunhaga 3, IS-107 Reykjavik, Iceland}
\author{Nzar Rauf Abdullah}
\affiliation{Physics Department, College of Science, 
             University of Sulaimani, Kurdistan Region, Iraq}
\affiliation{Komar Research Center, Komar University of Science and Technology, Sulaimani, Kurdistan Region, Iraq}
\author{Chi-Shung Tang}
\email{cstang@nuu.edu.tw}
\affiliation{Department of Mechanical Engineering, National United University, Miaoli 36003, Taiwan}
\author{Andrei Manolescu}
\email{manoles@ru.is}
\affiliation{School of Science and Engineering, Reykjavik University, Menntavegur 
             1, IS-101 Reykjavik, Iceland}
\author{Valeriu Moldoveanu}
\email{valim@infim.ro}
\affiliation{National Institute of Materials Physics, PO Box MG-7, Bucharest-Magurele, Romania}

%
%----------------------------------------------------------------

\begin{abstract}
We model theoretically the time-dependent transport through an asymmetric double quantum dot etched in 
a two-dimensional wire embedded in a FIR photon cavity. For the transient and the intermediate time regimes the 
currents and the average photon number are calculated by solving a Markovian master equation in the dressed-states 
picture, with the Coulomb interaction also being taken into account. We predict that in the presence of a 
transverse magnetic field the interdot Rabi oscillations appearing in the intermediate and transient 
regime coexist with slower non-equilibrium fluctuations in the occupation of states for opposite 
spin orientation. The interdot Rabi oscillation induces charge oscillations across the system and 
a phase difference between the transient source and drain currents. We point out a difference between 
the steady-state correlation functions in the Coulomb blocking and the photon-assisted transport regimes. 
\end{abstract}

\maketitle
%
%----------------------------------------------------------------------------------------
%

\section{Introduction}
Experimental\cite{PhysRevX.6.021014,Delbecq11:01,2017arXiv170401961L,PhysRevX.7.011030,Frey11:01,Mi}
and theoretical\cite{PhysRevLett.116.113601,PhysRevB.92.165403,Gudmundsson16:AdP_10,2017arXiv170300803H,PhysRevB.97.195442}
interest is growing in electron transport through semiconductor systems in photon cavities.
The successes of circuit QED devices with superconducting quantum bits coupled to microwave cavities
have pushed for the evolution of hybrid mesoscopic circuits combining nanoconductors and 
metallic reservoirs.\cite{0953-8984-29-43-433002} Eventually, this effort might lead to 
the evolution of devices active in the challenging Terahertz-regime opening up novel
possibilities.\cite{0953-8984-29-43-433002} This has lead us to consider aspects of the electron-photon     
interactions on quantum transport in the far-infrared (FIR) regime.

The time-dependent electronic transport through a 2D nanosystem patterned in a GaAs 
heterostructure which is in turn embedded in a 3D FIR photon cavity generally displays
three regimes: i) The switching transient regime in which electrons tunnel through the system
but their interactions with the photons have not had time to affect the transport yet. 
ii) The intermediate regime during which the electron-photon coupling plays an important
role in bringing the system to a steady state. iii) The stationary regime with coexisting
radiative transitions and photon-assisted tunneling.\cite{Gudmundsson16:AdP_10}  
The characteristic time designated to the transient and the intermediate regime depends on the 
the ratio of the system lead coupling and the electron-photon coupling in addition to the
shape or geometry of the central system and the photon energy.\cite{2016arXiv161109453G}

In earlier publications we have shown how a Rabi oscillation can be detected in the 
transport of electrons through an electronic system in a 3D photon cavity, in the 
transient regime directly from the charge current,\cite{doi:10.1021/acsphotonics.5b00115}
and in the steady state through the Fourier power spectrum of the current current 
correlation function.\cite{GUDMUNDSSON20181672}  
Here, we will analyze the intermediate time regime and show that oscillations of the 
transport current in time still reveal Rabi oscillations, but in a complex many level 
system other oscillations can be present. In particular we find that for a weak Rabi splitting
the still weaker Zeeman spin splitting caused by a small external magnetic field plays a role
in the transport, but only in this regime dominated by strong nonequilibrium processes. 

In the earlier calculations the central system was a short quantum wire with parallel 
quantum dots of same shape. The anisotropy of the system makes the first excitation of
the even parity one-electron ground state to be an odd parity state with respect to the 
axis of the quantum dots, the $y$-axis. Subsequently, $y$-polarized cavity photons 
couple the two states strongly through the paramagnetic electron-photon interaction, but
only weakly through the diamagnetic interaction. $x$-polarized photon, on the other hand,
can only couple the two states weakly through the diamagnetic interaction. We thus observed
two different Rabi-oscillations depending on the polarization of the cavity 
field.\cite{doi:10.1021/acsphotonics.5b00115,GUDMUNDSSON20181672,Gudmundsson16:AdP_10,ANDP:ANDP201500298}
Here, we select an asymmetric system with slightly dissimilar quantum dots located 
at opposite ends of the short quantum wire. Consequently, the energy levels of each dot are 
different (or misaligned). The dots are well separated, such that the charge probability density distribution
of the lowest one-electron energy states of each dot are almost entirely located in each dot.
The lower one of them is the one-electron ground state of the system, and the other one is the 
first excited one-electron state state. We select the photon
energy to establish a Rabi-resonance between these two states. It is bound to be weak
as it relies on the small charge overlap of the states, but it is also interesting
as it promotes a charge oscillation over the entire length of the short quantum wire.   

In Section II we present the model and the quantum master equation formalism, Section III
contains the numerical results and their discussion while Section IV is left to conclusions.

\section{System and Model}
\begin{figure}[htb]
      \centerline{\includegraphics[width=0.30\textwidth]{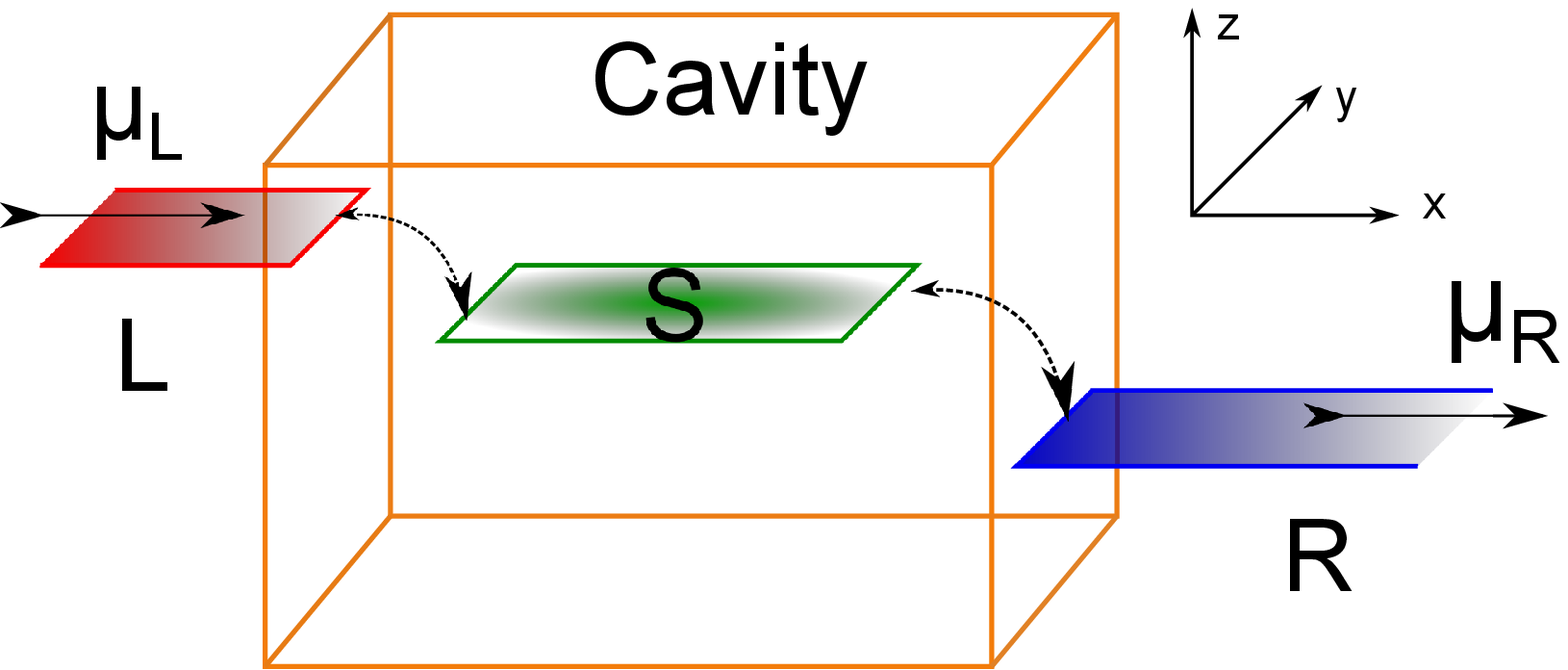}}
      \centerline{\includegraphics[width=0.42\textwidth]{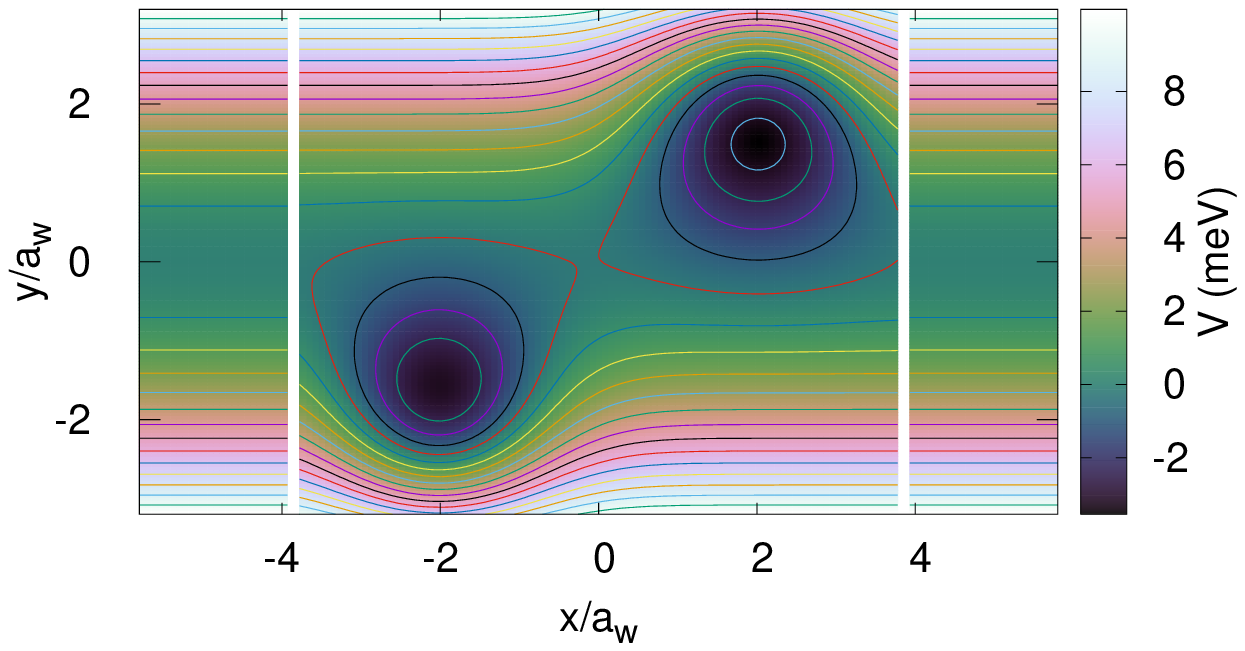}}
      \caption{(Upper) Schematic for a short quantum wire in a 3D photon cavity coupled to
               semi-infinite external left (L) and right (R) leads with quasi 1D electron systems
               with chemical potentials $\mu_L$ and $\mu_R$, respectively. 
               The electrons in the short quantum wire and the photons of the cavity
               comprise the central system (S).
               (Lower) The potential energy landscape defining the two asymmetrically quantum dots 
               embedded in a short quantum wire of length $L_x=180$ nm $\approx$ 7.6$a_w$,
               where $a_w=23.8$ nm is the effective magnetic length for transverse magnetic field
               $B_{\mathrm{ext}}=0.1$ T and parabolic confinement energy $\hbar\Omega_0=2.0$ meV of the 
               short wire and leads in the $y$-direction. The white gaps at $x\approx\pm{3.8}a_w$
               indicate the onset of the semi-infinite leads. The right dot is slightly deeper and 
               broader than the left dot.}
      \label{System}
\end{figure}

We consider a short quantum wire of length $L=180$ nm with two asymmetrically placed shallow 
quantum dots as is displayed in Fig.\ \ref{System}.   
The potential landscape defining the short quantum wire and dots is described
by
\begin{align}
      V(x,y) =& \left[\frac{1}{2}m^*\Omega_0^2y^2 +eV_g\right.\nonumber\\
             +& \left. \sum_{i=1}^2 V_d^i\exp{\left\{-\beta_i^2 (x-x_{0i})^2-\beta_i^2(y-y_{0i})^2\right\}} \right]\nonumber\\
             \times&\theta\left(\frac{L_x}{2}-|x|\right)
\label{Potential}
\end{align}
with $\hbar\Omega_0 = 2.0$ meV, $V_d^1 = -6.6$ meV, $V_d^2 = -6.8$ meV,
$\beta_1 = 0.030$ nm$^{-1}$, $\beta_2 = 0.028$ nm$^{-1}$, 
$x_{01}=-48$ nm, $x_{02}=+48$ nm, $y_{01}=-50$ nm, $y_{02}=+50$ nm,
$L_x = 180$ nm, and $\theta$ the Heaviside 
unit step function. The plunger gate voltage $V_g$ moves the states of the
system up or down in energy with respect to the bias window defined by the chemical potentials of the 
external leads to be describe below.

The Hamiltonian of the closed central system, the electrons and the photons, in terms of field operators is
\begin{align}
      H_\mathrm{S} =& \int d^2r \psi^\dagger (\mathbf{r})\left\{\frac{\pi^2}{2m^*}+
        V(\mathbf{r})\right\}\psi (\mathbf{r})
        + H_\mathrm{EM} + H_\mathrm{Coul}\nonumber\\ 
        -&\frac{1}{c}\int d^2r\;\mathbf{j}(\mathbf{r})\cdot\mathbf{A}_\gamma
        -\frac{e}{2m^*c^2}\int d^2r\;\rho(\mathbf{r}) A_\gamma^2,
\label{Hclosed}
\end{align}
with 
\begin{equation}
      {\bm{\pi}}=\left(\mathbf{p}+\frac{e}{c}\mathbf{A}_{\mathrm{ext}}\right).
\end{equation}
The static electron-electron Coulomb interaction is described by $H_\mathrm{Coul}$ with 
the kernel
\begin{equation}
      V_{\mathrm{Coul}}(\mathbf{r}-\mathbf{r}') = \frac{e^2}{\kappa_e\sqrt{|\mathbf{r}-\mathbf{r}'|^2+\eta_c^2}},
\label{VCoul}
\end{equation}
and a small regularizing parameter $\eta_c/a_w=3\times 10^{-7}$ ($a_w$ being defined below).
The second line of the Hamiltonian (\ref{Hclosed}) is the para- and the diamagnetic electron-photon
interactions, respectively. 
$\mathbf{A}_{\mathrm{ext}}$ is a classical vector potential leading to an homogeneous external
small magnetic field $B_{\mathrm{ext}}=0.1$ T directed along the $z$-axis, perpendicular the two-dimensional 
quantum wire, inserted to break the spin and possible orbital degeneracies of the states in order to enhance 
the stability of the results. We use GaAs parameters with $m^*=0.067m_e$, $\kappa_e=12.4$, and 
$g^*=-0.44$.  
The small external magnetic field, $B_{\mathrm{ext}}$, and the parabolic confinement energy of the leads and the 
central system $\hbar\Omega_0=2.0$ meV, together with the cyclotron frequency 
$\omega_c=(eB_{\mathrm{ext}})/(m^*c)$ produce an effective characteristic confinement energy 
$\hbar\Omega_w=\hbar({\omega_c^2+\Omega_0^2})^{1/2}$, and an effective magnetic length 
$a_w=(\hbar /(m^*\Omega_w))^{1/2}$.
This characteristic length scale is approximately 23.8 nm for the parameters selected here.
The Hamiltonian for the single cavity photon mode is 
$H_\mathrm{EM}=\hbar\omega a^\dagger a$, with energy $\hbar\omega$, in terms of the cavity photon 
annihilation and creation operators, $a^\dagger$ and $a$.
We assume a rectangular 3D photon cavity $(x,y,z)\in\{[-a_\mathrm{c}/2,a_\mathrm{c}/2]
\times [-a_\mathrm{c}/2,a_\mathrm{c}/2]\times [-d_\mathrm{c}/2,d_\mathrm{c}/2]\}$ 
with the short quantum wire located in the center of the $z=0$ plane. 

In the Coulomb gauge the polarization of the electric field of the cavity photons parallel to the transport
in the $x$-direction (with the unit vector $\mathbf{e}_x$) is realized in the TE$_{011}$ mode, 
or perpendicular to the transport (defined by the unit vector $\mathbf{e}_y$) in the 
TE$_{101}$ mode. 
The two modes of the quantized vector potential for the cavity field can be expressed as (in a stacked notation) 
\begin{equation}
      \mathbf{A}_\gamma (\mathbf{r})=\left({\hat{\mathbf{e}}_x \atop \hat{\mathbf{e}}_y}\right)
      {\cal A}\left\{a+a^{\dagger}\right\}
      \left({\cos{\left(\frac{\pi y}{a_\mathrm{c}}\right)}\atop\cos{\left(\frac{\pi x}{a_\mathrm{c}}\right)}} \right)
      \cos{\left(\frac{\pi z}{d_\mathrm{c}}\right)},
\label{Cav-A}
\end{equation}
with the strength of the vector potential, ${\cal A}$,
and the electron-photon coupling constant related by $g_\mathrm{EM} = e{\cal A}\Omega_wa_w/c$, 
leaving a dimensionless polarization tensor
\begin{equation}
      g_{ij}^k = \frac{a_w}{2\hbar}\left\{\langle i|\hat{\mathbf{e}}_k\cdot\bm{\pi}|j\rangle + \mathrm{h.c.}\right\},
\end{equation}
to be evaluated, where $k=x$ or $y$ is the polarization of the photon field.  
We will assume $g_\mathrm{EM}$ to have values of 0.05, or 0.10 meV here. As we are treating a many level system
with some transitions in resonance with the photon field and others not, we will not use the rotating wave
approximation for the electron-photon interactions in the central system.

The coupling of the central system to the leads, functioning as electron and energy reservoirs, 
is accomplished with the Hamiltonian
\begin{equation}
    H_T = \theta (t)\sum_{il} \int d  \mathbf{q} \left(T_{\mathbf{q}i}^l c_{\mathbf{q}l}^\dagger d_i +
          (T_{\mathbf{q}i}^l )^* d_i^\dagger c_{\mathbf{q}l}\right),
\label{H_T}
\end{equation} 
where $d_i$ is an annihilation operator for the single-electron state $|i\rangle$ of the 
central system, $c_{\mathbf{q}l}$ an annihilation operator for an electron in lead $l\in\{L,R\}$ 
in state $|{\mathbf q}\rangle$, with ${\mathbf q}$ standing for the momentum $q$ and the subband index
$n_l$ in the semi-infinite quasi-one dimensional lead. The coupling tensor $T_{\mathbf{q}i}^l$ depends
on the nonlocal overlap of the single-electron states at the internal boundaries in the central system and the respective 
lead.\cite{Gudmundsson09:113007,Moldoveanu09:073019,Gudmundsson12:1109.4728}
This approach describing a weak tunneling coupling of the central system and the leads
allows for full coupling between the quantum dots and the rest of the central system, like
in a scattering approach.\cite{Gudmundsson05:BT} Moreover, it conserves parity of states 
in the transition between the leads and the system. {All details and parameters of the coupling scheme
have been published in Eqs.\ (13-14) and the caption of Fig.\ 6 in Ref.\ \onlinecite{Gudmundsson12:1109.4728}}. 
The remaining overall coupling constant to the leads is $g_{\mathrm{LR}}a_w^{3/2}=0.124$ meV,
in the weak coupling limit used here.

We will investigate here the physical properties of the open system in the intermediate time
range where radiative transitions are active and touch upon the long time evolution. 
We thus revert to a description based on a Markovian version of a Nakajima-Zwanzig generalized master
equation\cite{Nakajima58:948,Zwanzig60:1338} that has been derived constructing the kernel of the 
integro-differential equation upto second order in the lead-system interaction (\ref{H_T}).

We assume a leaky photon cavity described by weakly coupling the single cavity photon mode via the 
lowest order dipole interaction to a reservoir of photons. For this interaction we assume a rotating
wave approximation. Care has to be taken in deriving the corresponding damping terms for the 
master equation as they have to be transformed from the basis of non-interacting photons to the 
basis of interacting electrons and photons (the eigenstates of $H_\mathrm{S}$ (\ref{Hclosed})).  
\cite{PhysRev.129.2342,PhysRevA.31.3761,PhysRevA.84.043832,PhysRevA.80.053810,PhysRevA.75.013811} 
In the Schr{\"o}dinger picture used here this can be performed by neglecting all creation terms in the 
transformed annihilation operators, and all annihilation terms in the transformed creation operators.
This guarantees that the open system will evolve into the correct physical steady state with respect to 
the photon decay. The Markovian master equation for the reduced density operator $\rho_\mathrm{S}$ has the form
\begin{align}
      \partial_t\rho_\mathrm{S}(t) = &-\frac{i}{\hbar}[H_\mathrm{S},\rho_\mathrm{S}(t)]
      -\left\{\Lambda^L[\rho_\mathrm{S} ;t]+\Lambda^R[\rho_\mathrm{S} ;t]\right\}\nonumber\\
      &-\frac{\kappa}{2\hbar}(\bar{n}_\textrm{R}+1)\left\{2\alpha\rho_\mathrm{S}\alpha^\dagger - \alpha^\dagger \alpha\rho_\mathrm{S} 
                                                                                 - \rho_\mathrm{S}\alpha^\dagger \alpha\right\}\nonumber\\
       &-\frac{\kappa}{2\hbar}(\bar{n}_\textrm{R})\left\{2\alpha^\dagger\rho_\mathrm{S}\alpha  - \alpha\alpha^\dagger\rho_\mathrm{S} 
                                                                                 - \rho_\mathrm{S}\alpha\alpha^\dagger\right\} ,
\label{NZ-eq}
\end{align}
where the creation and annihilation operators $\alpha^\dagger$ and $\alpha$ represent the original operators
in the non-interacting photon number basis, $a^\dagger$ and $a$, transformed to the interacting electron photon
basis $\{|\breve{\mu})\}$ using the rotating wave approximation. We select the photon decay constant as $\kappa=1.0\times10^{-5}$ meV, and 
$\bar{n}_\textrm{R} =0$ or 1.
The electron dissipation terms, $\Lambda^{L,R}[\rho_\mathrm{S};t]$, in the first line of Eq.\ (\ref{NZ-eq}) are complicated
functionals of the reduced density operator $\rho_\mathrm{S}$, and are explicitly
given in Refs.\ \onlinecite{GUDMUNDSSON20181672} and \onlinecite{2016arXiv161003223J}.    

We vectorize the Markovian master equation transforming it from the $N_\mathrm{F}$-dimensional many-body Fock space of 
interacting electrons and photons to a $N_\mathrm{F}^2$-dimensional Liouville space of transitions. The resulting first
order linear system of coupled differential equations is solved analytically,\cite{PhysRevB.81.155303}
and the solution is effectively evaluated at all needed points in time using parallel methods for linear algebra operations 
in FORTRAN or CUDA.\cite{2016arXiv161003223J} The reduced density operator is used to calculate mean values of relevant
physical quantities, and the R{\'e}niy-2 entropy of the central 
system\cite{2017arXiv170708946S,2018arXiv180608441B,2011arXiv1102.2098B}
\begin{equation}
      S=-k_B\ln{[Tr(\rho^2_\mathrm{S})]}.
\label{entropy}
\end{equation}
The trace operation in Eq.\ (\ref{entropy}) is independent of the basis carried out in, but we use the fully 
interacting basis $\{|\breve{\mu})\}$ as we do also for the Markovian master equation (\ref{NZ-eq}).

\section{Results}
\subsection{Many-body states and spectrum}
The many-body states of the central system are constructed in a step wise fashion in order to
maintain a high accuracy of the numerical results.\cite{Gudmundsson:2013.305}
Initially, a Fock space of non-interacting electrons is constructed
from $N_\mathrm{ses}=36$ accurate single-electron states (SES) keeping enough one-, two-, and three-electron states in order for the energy of
of the highest states for each electron number to surpass the bias window defined by the chemical potential in 
the leads by much. For the selected parameters the total number of states is 1228 many-electron states (MES). 
This many-body basis is then used to diagonalize 
the Coulomb interacting (\ref{VCoul}) electron system. Second, a basis is constructed as a tensor product
of the $N_\mathrm{mes}=120$ lowest in energy Coulomb interacting electron states and the 16 lowest photon number  
operator eigenstates. These are subsequently used to diagonalize the closed electron-photon interacting system
and creating the states $|\breve{\mu})$. 
Finally, the lowest $N_\mathrm{F}=120$ in energy of these cavity-photon dressed electron states are used for the transport calculation. 
The step wise construction parallels the step wise construction of Green functions for an interacting electron-photon system.

\begin{figure}[htb]
      \centerline{\includegraphics[width=0.48\textwidth]{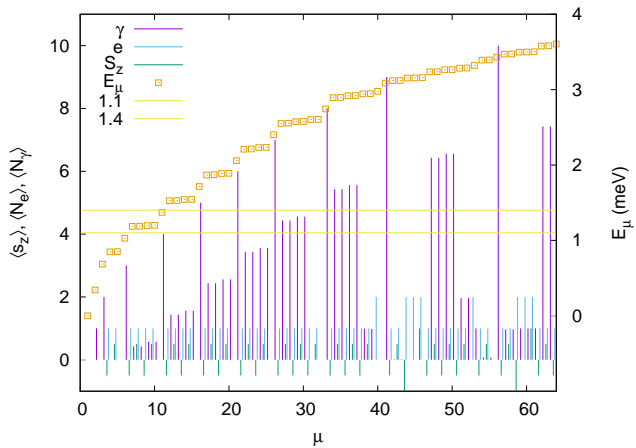}}
      \caption{The energy spectrum (golden squares) for the interacting closed central system for plunger gate voltage 
               $V_g=1.6$ mV and photon energy $\hbar\omega =0.343$ meV as a function of the state number $\mu$. 
               The electron, photon content, and $z$-component of spin is indicated with
               vertical bars for each many-body state. The chemical potentials of the left ($\mu_L$) and right ($\mu_R$) leads are 
               shown in relation with the spectrum. $B_{\mathrm{ext}}=0.1$ T, $L_x=180$ nm, $\hbar\Omega_0 =2.0$ meV, and $g_\textrm{EM}=0.05$ meV.
               $x$-polarized cavity photon field.}
      \label{NeNphESz-Vgp1p60-hw0p343-gEM0p10-x}
\end{figure}

The many-body energy spectrum, the electron and the mean photon content, and the $z$-component of the spin of the 
64 lowest in energy many-body states $|\breve{\mu})$ of the closed central system are displayed in 
Fig.\ \ref{NeNphESz-Vgp1p60-hw0p343-gEM0p10-x}.
The photon energy $\hbar\omega =0.343$ meV coupling the two lowest one-electron states mostly localized 
in each quantum dot leads to a Rabi-resonance showing up in non-integer values for the photon
content of some states. The probability density for both spin components of the one-electron ground state 
are almost entirely localized in the deeper quantum dot, the right dot, but due to the finite 
separation of the dots there is a very small probability for the electron to be found in the left dot.
The corresponding one-electron wavefunction has positive parity with respect to the dots.
These states, $|\breve{4})$ and $|\breve{5})$, have energies 0.8496 and 0.8521 meV, respectively,
and their first photon replicas, interact with the two spin components of
lowest energy one-electron state mostly localized in the left quantum dot. 
(The tiny overlap of the charge distribution between the dots causes the corresponding
single-electron wavefunction to have negative parity with respect to the dots).
The 4 resulting states 
$|\breve{7})$, $|\breve{8})$,  $|\breve{9})$, and $|\breve{10})$, all end up in 
the bias window defined by the chemical potentials of the left (L) and right (R) leads 
when the system is opened up for transport. Their energies as functions of the photon energy
$E_\textrm{EM}=\hbar\omega$  
\begin{figure}[htb]
      \includegraphics[width=0.48\textwidth]{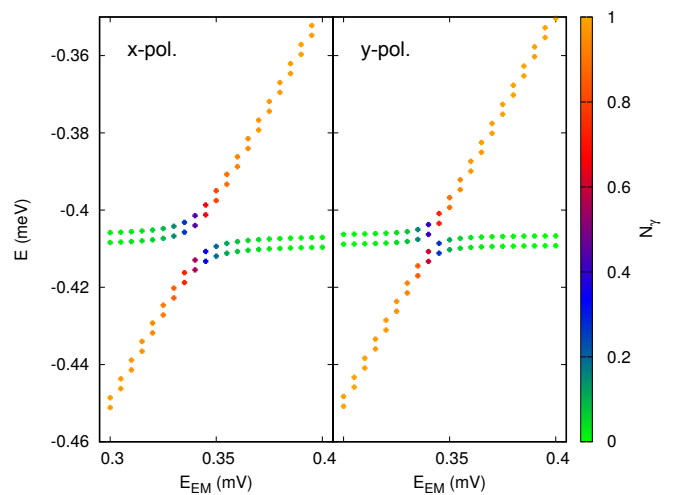}
      \caption{The energy of the two spin components of the lowest in energy one-electron states
               mostly localized in the left dot (shallower) and the first photon replica of the 
               corresponding states in the right dot (deeper)
               as the energy of the cavity photon $E_\textrm{EM}=\hbar\omega$ is varied.
               $V_g=0$ mV. For $V_g=1.6$ mV the states are labeled with  
               $|\breve{7})$, $|\breve{8})$, $|\breve{9})$, and $|\breve{10})$,
               see Fig.\ \ref{NeNphESz-Vgp1p60-hw0p343-gEM0p10-x}. 
               The mean photon content of the states is indicated by color of the dots and 
               defined by the color bar on the right side of the figure.
               (Left) $x$-polarized, and (right) $y$-polarized cavity photon field.
               $B_{\mathrm{ext}}=0.1$ T, $L_x=180$ nm, $\hbar\Omega_0 =2.0$ meV, and $g_\textrm{EM}=0.05$ meV.}
      \label{ERabi01-g010-xy-rof-AS-ColNph-hw}
\end{figure}
are shown in Fig.\ \ref{ERabi01-g010-xy-rof-AS-ColNph-hw} for a $x$-polarized cavity photon field
(left panel) and a $y$-polarized cavity photon field (right panel). The mean photon component of
the states and the anticrossings indicates a Rabi-splitting, that is a bit larger for the $x$-polarized cavity field as
the geometry of the system makes the charge densities of the states a bit more polarizable in that 
direction. As was mentioned earlier both Rabi splittings are small and not much larger than the 
Zeeman splitting of the states for $B=0.1$ T. Due to the weak charge overlap of the states almost
localized in each dot, and having opposite parity both the para- and the diamagnetic electron-photon
interactions contribute to the Rabi-resonance.   

\subsection{Transport}
\begin{figure}[htb]
      \includegraphics[width=0.45\textwidth]{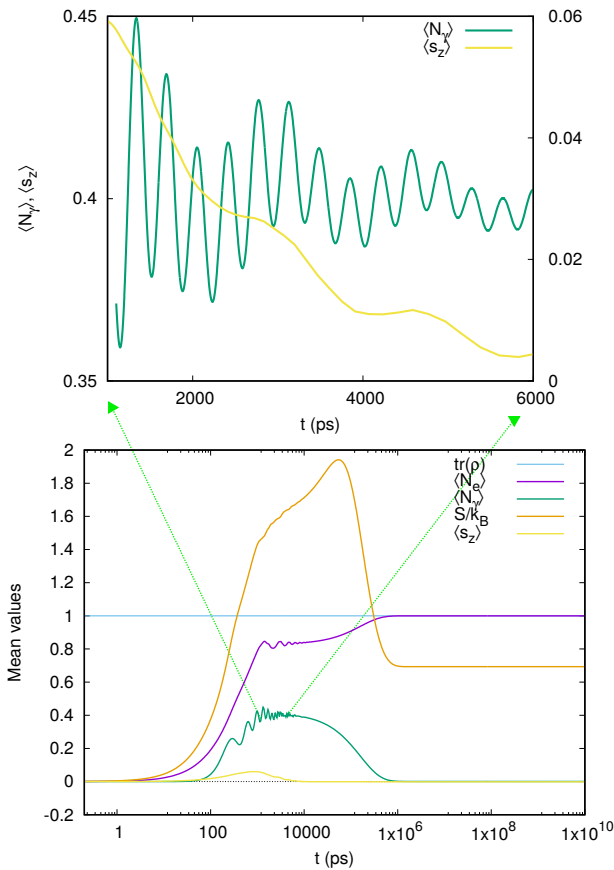}
      \caption{The mean electron $\langle N_e\rangle$ and photon $\langle N_\gamma\rangle$ content in the open central system as 
               functions of time. In addition are the mean $z$-component of the total spin, $\langle s_z\rangle$, trace of the reduced density
               matrix, and the entropy of the open central system $S/k_B$. The (upper panel) shows the details of $\langle N_\gamma\rangle$
               (left $y$-scale) and $\langle s_z\rangle$ (right $y$-scale) for the intermediate time range.
               $\hbar\omega =0.343$ meV, $\kappa=1.0\times10^{-5}$ meV, $B_{\mathrm{ext}}=0.1$ T, $g_\textrm{EM}=0.05$ meV, $\bar{n}_\mathrm{R}=0$,
               and $x$-polarized cavity photon.}
      \label{MeanV}
\end{figure}

As stated earlier, Rabi-oscillations in the transport current have been predicted in the transient 
regime,\cite{GUDMUNDSSON20181672} and current-current noise spectra in the steady state reveal 
their signs.\cite{doi:10.1021/acsphotonics.5b00115} Figure \ref{MeanV} displays the mean electron and 
photon numbers over the whole time scale (lower panel) relevant to the present model parameters for the case 
of initially empty central system.  
In addition the figure shows the mean value of the $z$-component of the total spin of the electrons, the trace 
of the reduced density matrix and the R{\'e}niy-2 entropy of the central system $S$ (\ref{entropy}). 
Initially, the central system gains electric charge through the states in the bias window. The plunger gate voltage
is placed at $V_g=1.6$ mV moving the one-electron ground state below the bias window. The steady state is reached
when the ground state is fully occupied and the system is Coulomb blocked with no mean current flowing through it. 
The entropy of the central system starts at zero as should be for an empty system. It rises in the intermediate
time range when many transitions are active in the central system, but does not return to zero in the steady state, which includes
both spin components of the one-electron ground state, and is thus not a pure state.

We notice that the mean photon number in the system only assumes a considerable value during the late charging
regime from 100 ps -- 0.6 ms, when radiative transitions assist in moving charge from the states in the bias window
to the ground state of the system.\cite{Gudmundsson16:AdP_10} 
The steady-state photon number vanishes because on one hand the cavity is lossy with $\kappa =1.0\times 10^{-5}$ meV 
and on the other hand the filling of the single-electron ground state prevents further radiative transitions.

We focus our attention on this intermediate regime, and
for a part of it we show the mean photon number and the $z$-component of the total spin in the upper panel of Fig.\ \ref{MeanV}.    
The mean photon number shows oscillations, a faster one that corresponds to the small Rabi splitting energy 
visible in the left panel of Fig.\ \ref{ERabi01-g010-xy-rof-AS-ColNph-hw}, and a slower oscillation that is 
also present in $\langle s_z\rangle$. This slower oscillations correlates with the effective Zeeman energy
$E_z=0.00255$ meV at $B_{\mathrm{ext}}=0.1$ T, corresponding to the period $T_z = 1624$ ps. The energy of the cavity photon,
$\hbar\omega =0.343$ meV corresponds to the time period $T=12.1$ ps, and is not seen in Fig.\ \ref{MeanV}.  

\begin{figure}[htb]
      \includegraphics[width=0.42\textwidth]{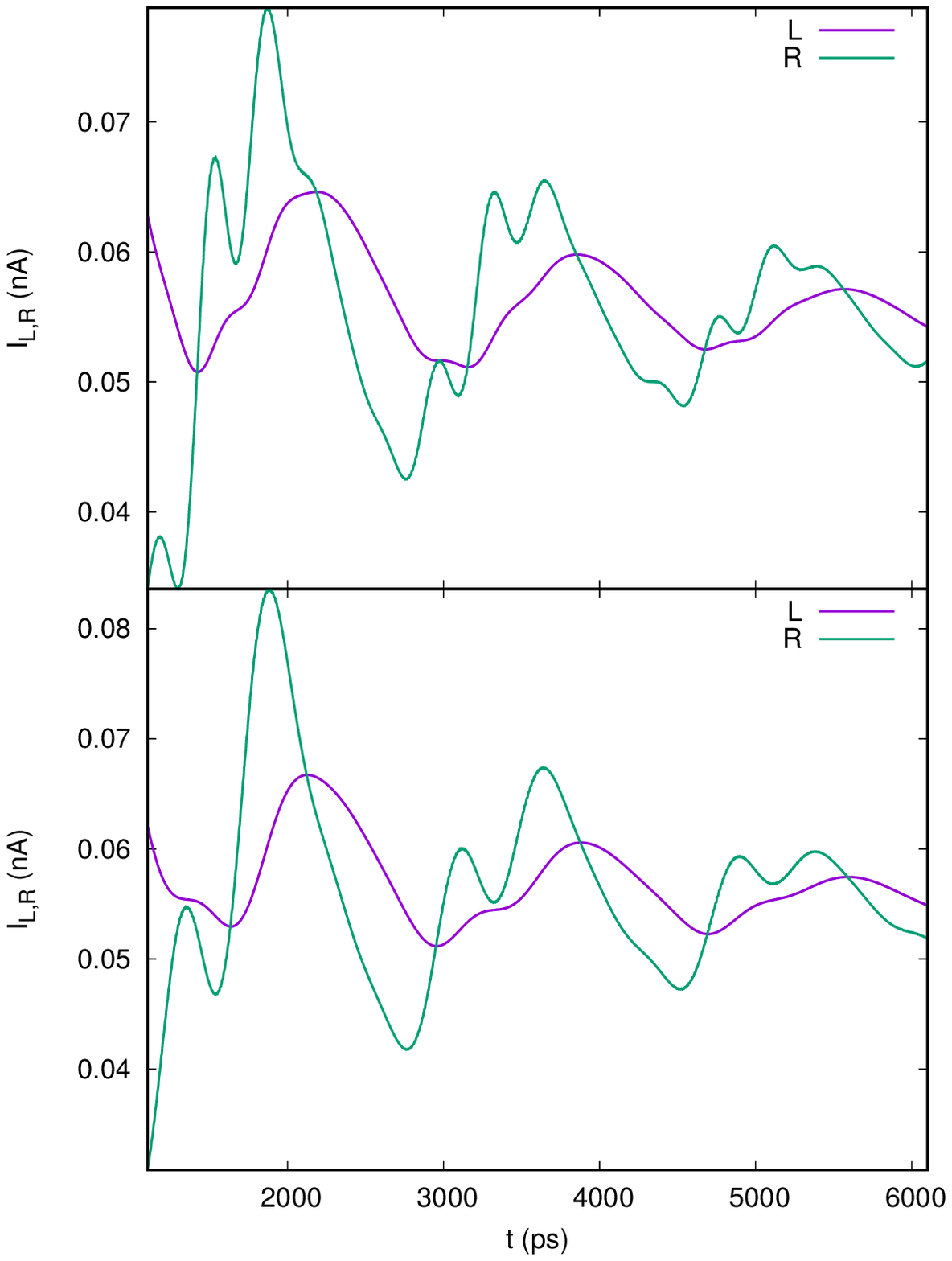}
      \caption{The current from the left lead (L) into the central system and the current from the central
               system into the right lead (R), for $x$- (Upper), and $y$-polarized (lower) cavity photon field.
               $g_\textrm{EM}=0.05$ meV, $\hbar\omega =0.343$ meV, $\kappa=1.0\times10^{-5}$ meV, $\bar{n}_\mathrm{R}=0$,
               and $B_{\mathrm{ext}}=0.1$ T. }
      \label{Current-10}
\end{figure}

In order to confirm this identification of oscillations we analyze the left current, $I_L$, into the central
system and the right current, $I_R$, out of it. The transport current can give us further insight into 
the dynamics in the system. It is displayed in Fig.\ \ref{Current-10} for the same parameters as
were used in Fig.\ \ref{MeanV}.
In the upper panel are the currents for the $x$-polarized cavity photon field and for the 
$y$-polarized one in the lower panel. The right dot is slightly deeper and wider and its states
should have a slightly better coupling to the right lead and the states in the left dot to the 
left lead. The one-electron ground state is mostly localized in the right dot with its first 
photon replica in the bias window. Fig.\ \ref{Current-10} shows clear Rabi-oscillations in $I_R$
and much weaker in $I_L$. The Rabi-oscillations are a bit faster for the $x$-polarized photon field
than the $y$-polarized in accordance with the Rabi-energies readable 
from the anticrossing levels in Fig.\ \ref{ERabi01-g010-xy-rof-AS-ColNph-hw}. 
Additionally, we notice what seems to be an offset or a phase difference between the 
left and right current. We address this issue below. 

\begin{figure}[htb]
      \includegraphics[width=0.42\textwidth]{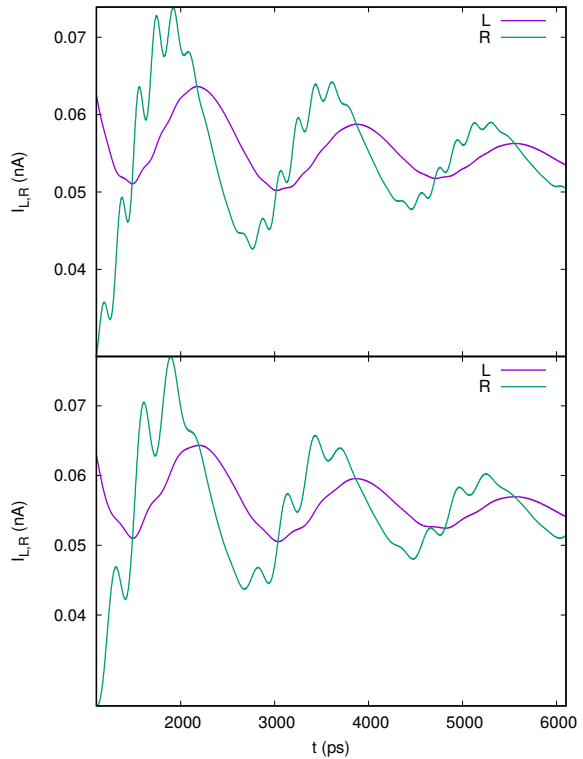}
      \caption{The current from the left lead (L) into the central system and the current from the central
               system into the right lead (R), for $x$- (Upper), and $y$-polarized (lower) cavity photon field.
               $g_\textrm{EM}=0.10$ meV, $\hbar\omega =0.343$ meV, $\kappa=1.0\times10^{-5}$ meV, $\bar{n}_\mathrm{R}=0$,
               and $B_{\mathrm{ext}}=0.1$ T.}
      \label{Current-20}
\end{figure}

First, we observe the transport currents for a a higher electron-photon coupling in Fig.\ \ref{Current-20},
where $g_\mathrm{EM}=0.1$ meV, instead of 0.05 meV in Fig.\ \ref{Current-10}. 
We notice that the Rabi frequency doubles, like should be expected, for both polarization of the cavity field,
but the frequency of the slower oscillations is not changed. 

\begin{figure}[htb]
      \includegraphics[width=0.42\textwidth]{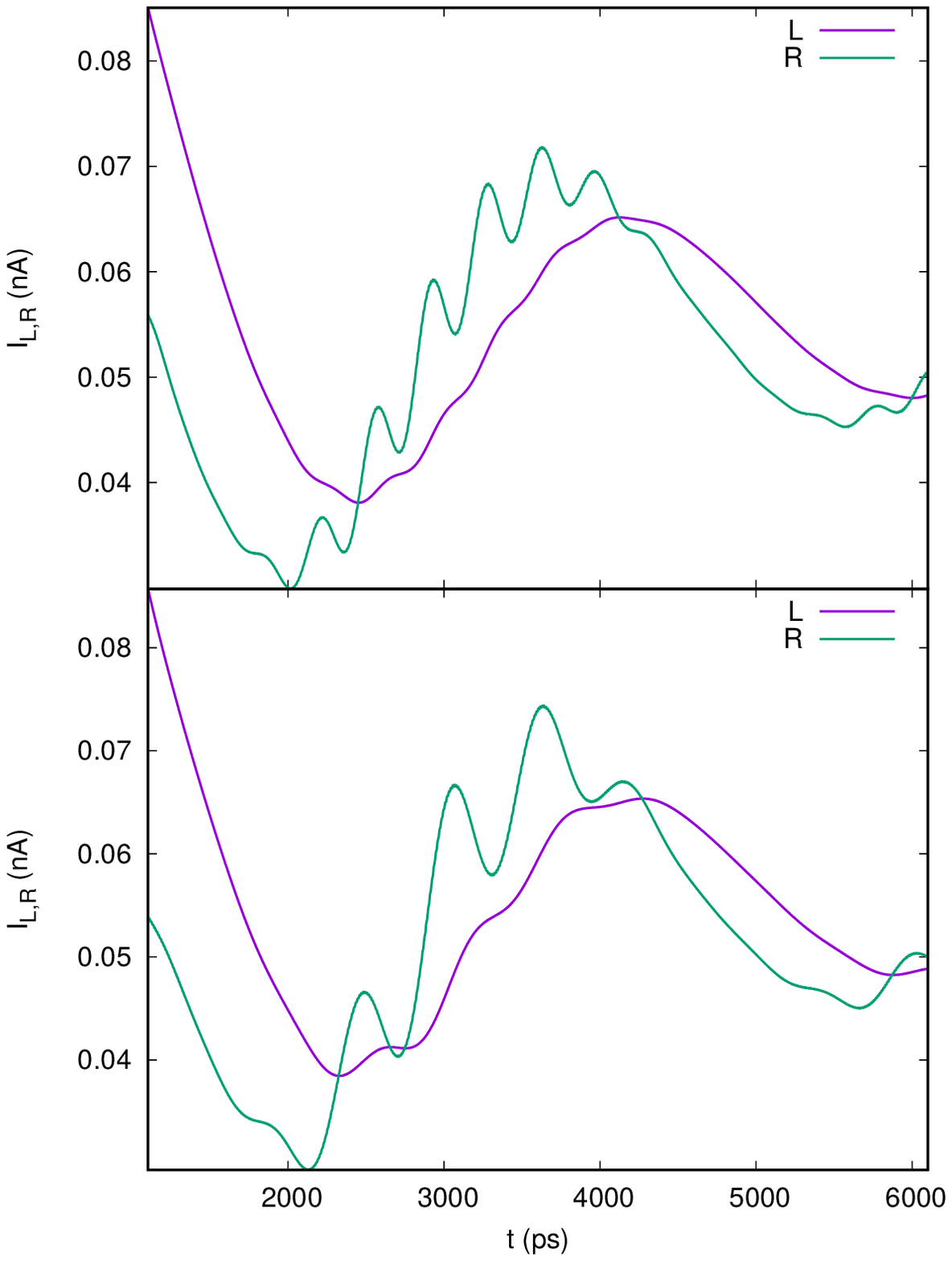}
      \caption{The current from the left lead (L) into the central system and the current from the central
               system into the right lead (R), for $x$- (Upper), and $y$-polarized (lower) cavity photon field.
               $g_\textrm{EM}=0.05$ meV, $\hbar\omega =0.343$ meV, $\kappa=1.0\times10^{-5}$ meV, $\bar{n}_\mathrm{R}=0$,
               and $B_{\mathrm{ext}}=0.05$ T. }
      \label{Current-10-B0p05}
\end{figure}

If the slower oscillations are linked to the Zeeman splitting then their frequency should change with the 
small external magnetic field perpendicular to the short quantum wire. In Fig.\ \ref{Current-10-B0p05} we
keep the electron-photon coupling $g_\mathrm{EM}=0.05$ meV, but reduce $B_{\mathrm{ext}}$ from 0.1 T to 0.05 T.
Indeed, the period of the slower oscillation doubles and the faster oscillation remains constant.

\begin{figure}[htb]
      \includegraphics[width=0.45\textwidth]{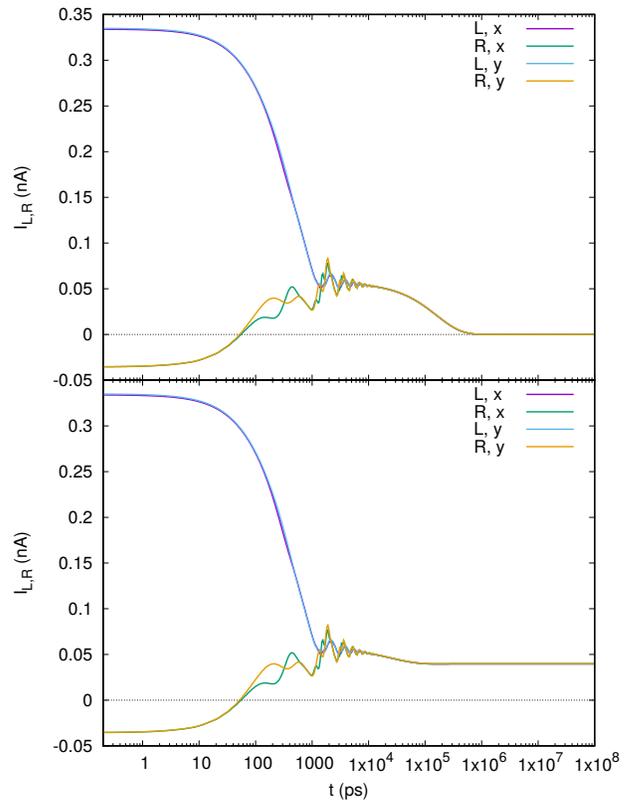}
      \caption{The current from the left lead (L) into the central system and the current from the central
               system into the right lead (R) for $x$- and $y$-polarized cavity photon field for the 
               whole time range from the transient to the steady state regime. The constant average photon 
               number in the reservoir $\bar{n}_\textrm{R}=0$ (upper), and  $\bar{n}_\textrm{R}=1$ (lower).
               $g_\textrm{EM}=0.05$ meV, $\hbar\omega =0.343$ meV, $\kappa=1.0\times10^{-5}$ meV, $B_{\mathrm{ext}}=0.1$ T. }
      \label{Current-At}
\end{figure}

In Fig.\ \ref{Current-At} we show the currents for the whole time scale. In the upper panel 
we have selected, as above, the photon reservoir to be empty, $\bar{n}_\mathrm{R} =0$. 
In this case the system is charged and enters ultimately a Coulomb-blocked steady state with no 
transport current.
In the lower panel of Fig.\ \ref{Current-At} we assume $\bar{n}_\mathrm{R} =1$, and in the 
steady state we have a photon assisted transport. In this case (not shown here) the entropy $S$ is
not reduced as the system enters the steady state as all photon active transitions remain active.
Clearly seen in Fig.\ \ref{Current-At} is the phase difference between the left and right transport 
current, even though the logarithmic time scale washes this effect out.

\begin{figure}[htb]
      \includegraphics[width=0.48\textwidth]{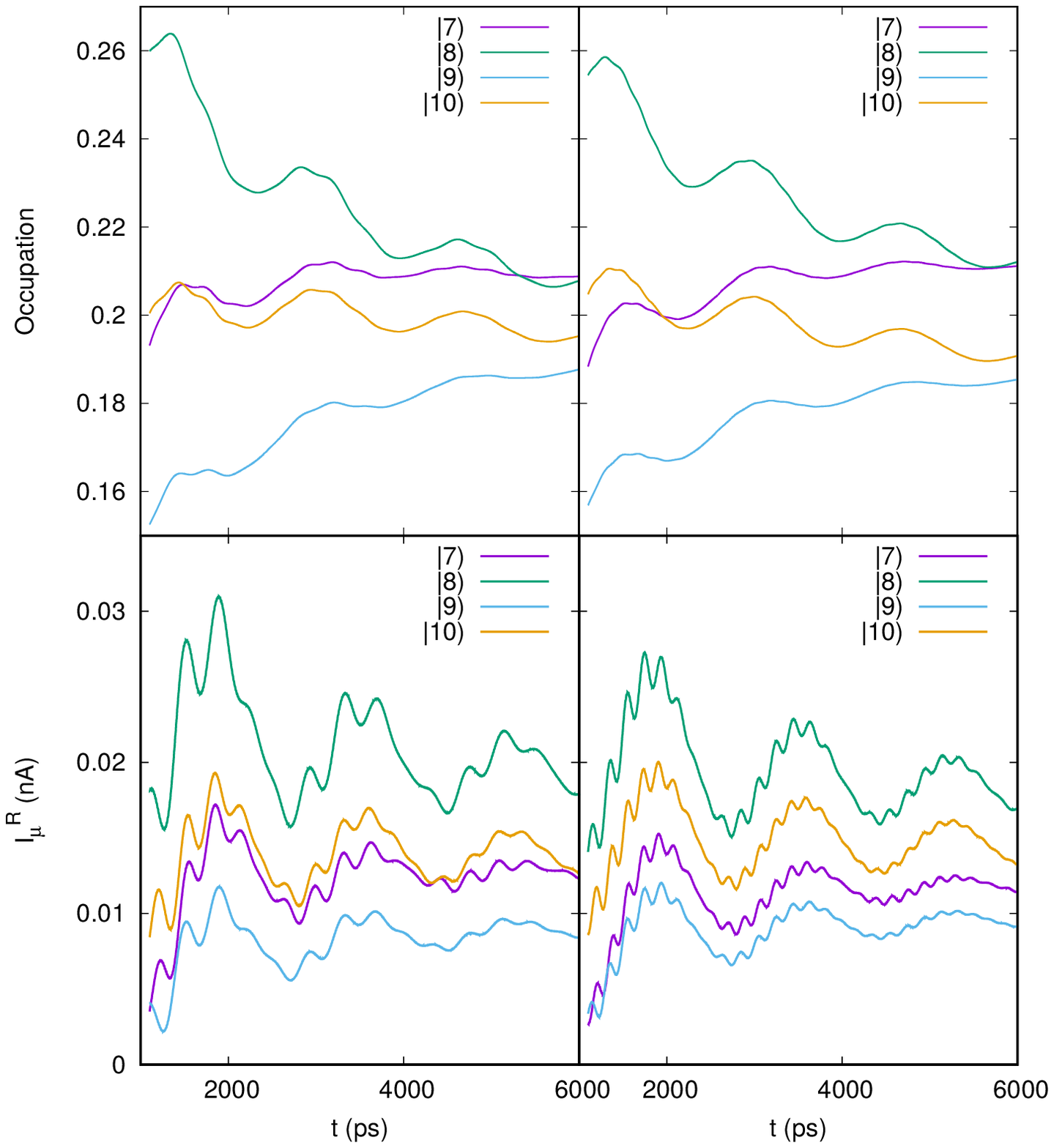}
      \caption{(Upper) The time dependent partial current through the photon dressed one-electron states 
               $|\breve{7})$, $|\breve{8})$, $|\breve{9})$, and $|\breve{10})$
               in the bias window for electron-photon coupling $g_\textrm{EM}=0.05$ meV (left),
               and $g_\textrm{EM}=0.10$ (right).  
               The partial current from the same states from the central system to the left
               lead for electron-photon coupling $g_\textrm{EM}=0.05$ meV (left),
               and $g_\textrm{EM}=0.10$ (right).
               According to Fig.\ \ref{NeNphESz-Vgp1p60-hw0p343-gEM0p10-x} states $|\breve{7})$
               and $|\breve{9})$ have spin quantum number $s_z=-1/2$, but states $|\breve{8})$
               and $|\breve{10})$ have $s_z=+1/2$.
               $\kappa=1.0\times10^{-5}$ meV, $B_{\mathrm{ext}}=0.1$ T, and $x$-polarized cavity photons.}
      \label{Corr-x}
\end{figure}

To investigate the reasons for the oscillations with the Zeeman energy we analyze the occupation 
of the states in the bias window active in the transport in the intermediate time range in the upper panels 
of Fig.\ \ref{Corr-x} for the case of a $x$-polarized cavity field, and two values of the 
electron-photon coupling, $g_\textrm{EM}=0.05$ meV (left panel) and $g_\textrm{EM}=0.1$ meV
(right panel).
The lower two panels of Fig.\ \ref{Corr-x} show the partial current through the same states, also for 
the two different values of $g_\mathrm{EM}$. We come back to this information below. 
The partial currents and occupation information are not experimental quantities, but they give us insight
into the dynamics in the system. We remember, as is seen in Fig.\ \ref{NeNphESz-Vgp1p60-hw0p343-gEM0p10-x}
that $|\breve{7})$ and $|\breve{8})$ have opposite $z$-components of the spin as do also states
$|\breve{9})$ and $|\breve{10})$, respectively, and we have no spin-orbit interaction in the system.
In the upper panels of the figure (Fig.\ \ref{Corr-x}) we see crossings of the occupation of states with 
opposite spin. Here, we have to have in mind that in the intermediate time regime the central system 
is in nonequilibrium and will evolve to a steady state with much more intuitive occupation distribution.
Moreover, the coupling to the leads of individual many-body states depends on the coupling of their 
single-particle components, their probability distribution in the contact area of the short quantum wire,
and depends on their energy and the density of states of the leads at the corresponding energy. 
The leads are quasi-1D with a sharply peaked density of states near the subband bottoms. 
Orbital magnetic effects are included in the leads, but their small Zeeman energy is 
neglected.\cite{Gudmundsson09:113007,Gudmundsson:2013.305}
With all this in mind it is clear that even the coupling of two spin components of the same state
to a state in a lead can be different, and the variable occupation of spin levels together with the tiny spin fluctuation
seen in Fig.\ \ref{MeanV} during the fastest changes in the system are nonequilibrium fluctuations.
Similar can be stated for the partial currents shown in the lower panels of Fig.\ \ref{Corr-x}.

The Rabi resonance for the photon energy $\hbar\omega =0.343$ meV entangles the lowest energy one-electron 
states that are mostly localized in each quantum dot. The time-dependent many-body charge distribution, or electron 
probability distribution, is thus expected to oscillate between the dots. In Video \ref{video} we see first the 
density at time $t=1102$ ps, and a click on the video icon produces a video with 100 frames equally spaced
for the time interval $t=1102$ -- $6000$ ps. 
\begin{video}
      \includegraphics[width=0.36\textwidth,bb=115 57 355 265,clip]{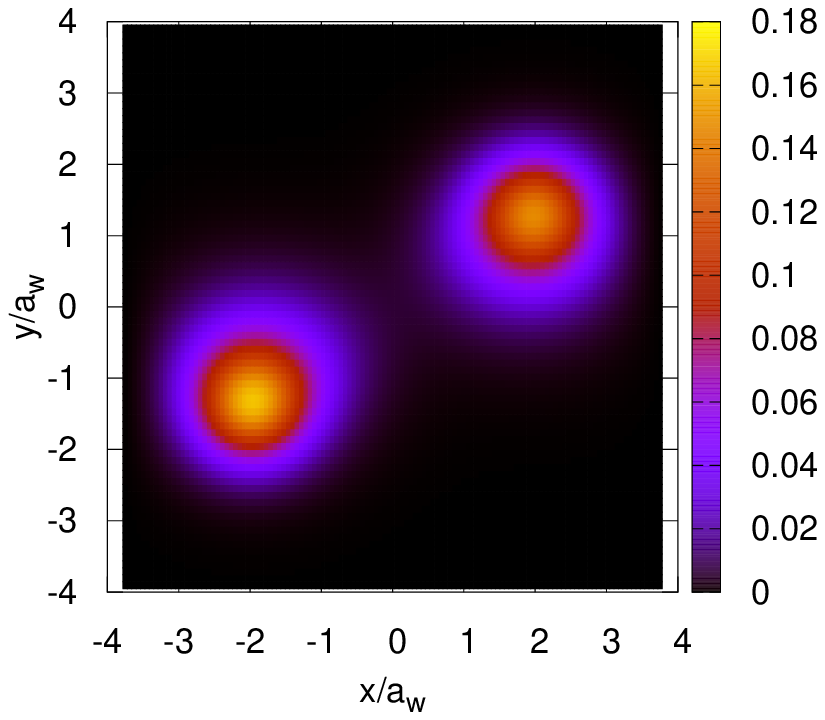}
%     \setfloatlink{Fig10.mp4}
%     \setfloatlink{Fig10.gif}
      \setfloatlink{https://youtu.be/oclm5LMpG-U}
      \caption{The many-body electron probability density at $t=1102$ ps for an initially empty
               central system. The time evolution till $t=6000$ ps in 100 steps can be seen
               by clicking on the label of this caption.
               $\mu_L=1.4$ meV, $\mu_R=1.1$ meV, $\hbar\omega =0.343$ meV, $\kappa=1.0\times10^{-5}$ meV, $B_{\mathrm{ext}}=0.1$ T.  }
      \label{video}
\end{video}
The video shows oscillations in the charge density between the dots with a combination of 
the Rabi- and the Zeeman frequency. This is in accordance with the left and right transport currents
displayed in Fig.\ \ref{Current-10}. Moreover, the charge oscillations in the video explain the
phase difference between the left and the right transport currents.

\begin{figure}[htb]
      \includegraphics[width=0.42\textwidth]{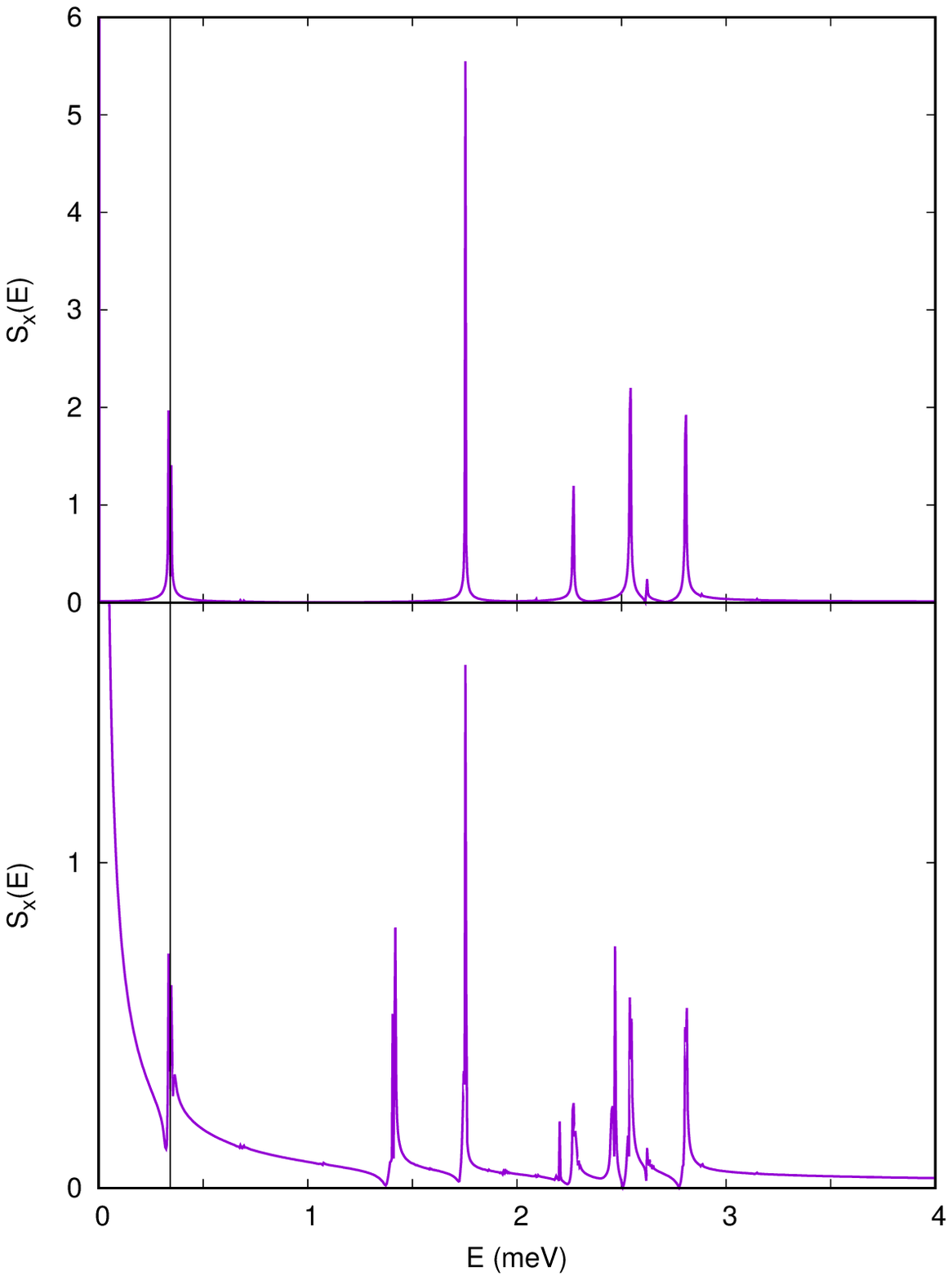}
      \caption{The Fourier power spectrum of the correlation function 
               $S_x(\tau )=\langle x(\tau)x(0)\rangle$ in the steady state.
               The cavity photon energy $\hbar$ is indicated by a vertical 
               line at 0.343 meV. The constant average photon 
               number in the reservoir $\bar{n}_\textrm{R}=0$ (upper), and  $\bar{n}_\textrm{R}=1$ (lower).
               $g_\textrm{EM}=0.05$ meV, $\hbar\omega =0.343$ meV, $\kappa=1.0\times10^{-5}$ meV, $B_{\mathrm{ext}}=0.1$ T. 
               The spectra are calculated using 5000 time points in the steady state regime.}
      \label{xy-Corr}
\end{figure}

Besides, the oscillations between the dots the Video \ref{video} indicates that there might be present
tiny faster oscillations of the density within each dot. This is not easy to quantify well within the finite 
intermediate time range, but can be investigated in the steady state using the correlation function
$S_x(\tau )=\langle x(\tau)x(0)\rangle$. In Fig.\ \ref{xy-Corr} we show the Fourier power spectrum of the 
correlation function $S_x(\tau )$ for $\bar{n}_\textrm{R}=0$ in the upper panel, and for $\bar{n}_\textrm{R}=1$ in the lower one. 
The spectra are calculated using the quantum regression theorem,\cite{0305-4470-14-10-013,Wallis-QO}
valid in the Markovian limit for weakly coupled systems. \cite{doi:10.1063/1.3570581,PhysRevA.47.1652,PhysRevB.59.10748,PhysRevB.64.235307,PhysRevB.92.165403}
Both spectra show a peak at the photon frequency $\hbar\omega =0.343$ meV and several higher energy peaks
that can be assigned to many-body transitions available in the system. 
The main peak at 1.75 meV is caused by a photon active transition between $|\breve{4})$ and $|\breve{5})$ to
the states $|\breve{31})$ and $|\breve{32})$ which are the first excitation of both spin components of the 
one-electron ground state in the right dot. This excitation is thus between states mainly localized in the 
right dot. The three transitions at 2.2688, 2.54, and 2.808 meV in the upper panel are not photon active 
transitions to higher states. The additional transitions in the lower panel are mostly additional photon active
transitions promoted by the presence of photons in the system.

For $\bar{n}_\textrm{R}=0$ the system enters
a Coulomb-blocked steady state, but for $\bar{n}_\textrm{R}=1$ a photon assisted transport current flows through it
like is seen in Fig.\ \ref{Current-At}. Interestingly, ``pink'' or white noise is seen in $S_x$ when current flows
through the system. Same type of noise is seen in Fourier power spectra of the current-current correlation
functions, not shown here. The occurrence of pink or white noise is well know in electronic systems and is here
probably caused by a multitude of active transitions for the open multi-level system. 

\section{Discussion}
In summary, we have modeled a nanoscale electron system of two slightly different quantum dots
that shows interdot Rabi-oscillations between the two lowest energy levels. In the intermediate
transient time regime the Rabi-oscillations lead to a phase difference in the currents out of, and into, 
the central system. Due to the fast changes in level populations in this regime through
radiative and nonradiative transitions we observe a coexisting spin oscillation even though
the electron-photon interactions conserve spin. 

The coexisting of the Rabi- and the Zeeman oscillations for the intermediate time range is not 
unique to the present system structure. It is also seen in a system of two parallel quantum dots 
embedded in a short quantum wire.\cite{2016arXiv161109453G,doi:10.1002/andp.201700334}
The main difference between these two cases is the strength of the ``interaction'' of the quantum
dots, or the overlap of the charge densities of the localized states in each quantum dot. 
For the case of the parallel quantum dots the charge overlap is large to the extent that no
eigenstates are localized in either dot, and one might view the system as one highly geometrically
anisotropic quantum dot. In that case quantum selection rules make the Rabi-resonance between the 
lowest lying one-electron states to be caused by the paramagnetic electron-photon interaction for 
$y$-polarized cavity field, and by the diamagnetic interaction for the $x$-polarization. 
So, the polarization can be used to change between strong or weak Rabi-resonance. Here, that is not
the case, the distance between the dots makes the Rabi-resonance rather weak for both photon
polarizations. 

The master equation used in the model is derived assuming weak coupling of the leads to the 
central system, to the effect that the kernel of the integro-differential equation is constructed
with the system-lead coupling (\ref{H_T}) up to second order. Are we sure this is not producing the 
oscillations of the occupation of the spin levels? Probably, we can never be completely sure, but
as the time scale for the system needed to attain the steady state shows, we are using a very weak
coupling. We have weakened the coupling further, within what is doable as the time scale then gets 
still longer producing strain on numerical accuracy, but still we see the corresponding 
spin oscillations. Another indicator is in the paragraph here above, the different effective strength
of the electron-photon interaction, as seen in the different strength for the Rabi-splitting, does
not affect the spin-oscillations in time, neither does a direct change in the strength of the 
electron-photon interaction, $g_\mathrm{EM}$. 

The important message we want to convey from our modeling of time-dependent electron transport through 
multilevel interacting nano scale two-dimensional semiconductor system embedded in 3D photon cavities 
is that the Rabi-oscillations in the central system can be detected in the transport current through 
them in all the time regimes characteristic for the corresponding system. Additionally, we observe that
the noise spectrum in the steady state depends on whether the system is really open for transport or is in 
a Coulomb blocking regime. 

The challenging Terahertz or FIR regime for semiconducting QED circuits offers interesting
possibilities for fundamental research into the electron-photon interactions and devices 
with new potentials.

\begin{acknowledgments}
This work was financially supported by the Research Fund of the University of Iceland,
the Icelandic Research Fund, grant no.\ 163082-051, 
and the Icelandic Infrastructure Fund. CST acknowledges support from Ministry of Science and
Technology of Taiwan under grant No.\ 106-2112-M-239-001-MY3. V.M.\ acknowledges financial 
support by the CNCS-UEFISCDI Grant PN-III-P4-ID-PCE-2016-0084.
\end{acknowledgments}

%
%----------------------------------------------------------------------------------------
%
%\balance
\frenchspacing
\bibliographystyle{apsrev4-1}
%\bibliography{mod_qd}
%

%
%
%----------------------------------------------------------------------------------------
%
\end{document}